\documentclass[3p,times,procedia]{elsarticle}
\usepackage{nupha_ecrc}

\volume{00}

\firstpage{1}

\journalname{Nuclear Physics A}

\runauth{P. Moreau et al.}

\jid{nupha}

\jnltitlelogo{Nuclear Physics A}

\usepackage{amssymb}

\usepackage[figuresright]{rotating}

\begin{document}

\begin{frontmatter}



\dochead{XXVIth International Conference on Ultrarelativistic Nucleus-Nucleus Collisions\\ (Quark Matter 2017)}

\title{Evidence for chiral symmetry restoration in heavy-ion collisions}


\author[FIAS,ITP,GSI]{P. Moreau}

\author[Giessen]{A. Palmese}
\author[Giessen]{W. Cassing}
\author[Giessen]{E. Seifert}
\author[Giessen]{T. Steinert}
\author[FIAS,ITP,GSI]{E. L. Bratkovskaya}

\address[FIAS]{Frankfurt Institute for Advanced Studies, Johann Wolfgang Goethe Universit\"{a}t, Frankfurt am Main, Germany}
\address[ITP]{Institute for Theoretical Physics, Johann Wolfgang Goethe Universit\"{a}t, Frankfurt am Main, Germany}
\address[GSI]{GSI Helmholtzzentrum f\"{u}r Schwerionenforschung GmbH, Darmstadt, Germany}
\address[Giessen]{Institut f\"{u}r Theoretische Physik, Universit\"{a}t Giessen, Germany}

\begin{abstract}
We study the effect of the chiral symmetry restoration (CSR) on heavy-ion collisions observables in the energy range $\sqrt{s_{NN}}$ = 3-20\,GeV within the Parton-Hadron-String Dynamics (PHSD) transport approach. The PHSD includes the deconfinement phase transition as well as essential aspects of CSR in the dense and hot hadronic medium, which are incorporated in the Schwinger mechanism for particle production. Our systematic studies show that chiral symmetry restoration plays a crucial role in the description of heavy-ion collisions at $\sqrt{s_{NN}}$ = 3-20\,GeV, realizing an increase of the hadronic particle production in the strangeness sector with respect to the non-strange one. Our results provide a microscopic explanation for the “horn” structure in the excitation function of the $K^+/\pi^+$ ratio: the CSR in the hadronic phase produces the steep increase of this particle ratio up to $\sqrt{s_{NN}} \approx $ 7 GeV, while the drop at higher energies is associated to the appearance of a deconfined partonic medium. Furthermore, the appearance/disappearance of the ‘horn’ structure is investigated as a function of the system size. We additionally present an analysis of strangeness production in the ($T,\mu_B$)-plane (as extracted from the PHSD for central Au+Au collisions) and discuss the perspectives to identify a possible critical point in the phase diagram.
\end{abstract}


\end{frontmatter}


\section{Introduction}
\label{Introduction}

\label{intro} In this contribution we summarize the results from our study in Ref. \cite{Palmese:2016rtq} that investigates the strangeness enhancement in nucleus-nucleus collisions \cite{rafelski,stock} or the 'horn' in the $K^+/\pi^+$
ratio \cite{GG99,GGS11}. Previously both phenomena have been addressed to a deconfinement transition. Indeed, the actual
experimental observation could not be described within conventional hadronic transport theory \cite{Jgeiss,Brat04,Weber} and remained a major challenge for microscopic approaches. Only recently, the Parton-Hadron-String Dynamics (PHSD), a transport approach describing heavy-ion collisions on the basis of partonic, hadronic and string degrees-of-freedom, obtained a striking improvement on this issue when including chiral symmetry restoration (CSR) in the string decay for hadronic particle production \cite{Cassing:2015owa}.

\section{Chiral symmetry restoration in the string fragmentation process}
\label{CSR}

Our studies are performed within the PHSD transport approach that
has been described in Refs. \cite{PHSD,PHSDrhic}. PHSD incorporates
explicit partonic degrees-of-freedom in terms of strongly
interacting quasiparticles (quarks and gluons) in line with an
equation-of-state from lattice QCD (lQCD) as well as dynamical
hadronization and hadronic elastic and inelastic collisions in the
final reaction phase.

\subsection{Strings in (P)HSD}
The string formation and decay represents the dominant particle production mechanism in nucleus-nucleus collisions for bombarding energies from $2\,$AGeV to $160\,$AGeV. In PHSD, the primary hard scatterings between nucleons are described by string formation and decay in the FRITIOF Lund model \cite{FRITIOF}. The production probability $P$ of massive $s\bar{s}$ or
$qq\bar{q}\bar{q}$ pairs is suppressed in comparison to light flavor production ($u\bar{u}$, $d\bar{d}$) according to the Schwinger-like formula \cite{Schwinger}, i.e.
\begin{equation}
	{P(s\bar{s}) \over P(u\bar{u})} ={P(s\bar{s}) \over P(d\bar{d})} = \gamma_s = \exp\left(-\pi
	{m_s^2-m_q^2\over 2\kappa} \right) ,
	\label{schwinger}
\end{equation}
with $\kappa\approx 0.176$~GeV$^2$ denoting the string tension
and $m_s, m_q=m_u=m_d$ the appropriate (dressed) strange and light quark masses. 
Inserting the constituent (dressed) quark masses $m^v_u \approx 0.33$~GeV
and $m^v_s \approx 0.5$ GeV in the vacuum a value of $\gamma_s \approx 0.3$ is
obtained from Eq.(1). This ratio is expected to be different in a nuclear medium and actually should depend on the in-medium quark condensate $<\bar{q}q>$.

\subsection{The scalar quark condensate}
As it is well known the scalar quark condensate $<\bar{q}q>$ is viewed as an order
parameter for the restoration of chiral symmetry at high baryon
density and temperature. In leading order the scalar quark condensate $\langle \bar q q
\rangle$ can be evaluated in a dynamical calculation as follows \cite{Toneev98}
\begin{equation}
	\frac{<\bar{q}q>}{<\bar{q}q>_V} = 1 - \frac{\Sigma_\pi}{f_\pi^2
		m_\pi^2}\rho_S - \sum\limits_h{\sigma_h \rho_S^h \over f_\pi^2
		m_\pi^2}, \label{condens2} \end{equation} 
where $\sigma_h$ denotes the $\sigma$-commutator of the relevant mesons $h$. 
Furthermore, $<\bar{q}q>_V$ denotes the vacuum condensate, $\Sigma_\pi \approx$
45 MeV is the pion-nucleon $\Sigma$-term, $f_\pi$ and $m_\pi$ the
pion decay constant and pion mass, respectively. The quantities $\rho_S$ and
$\rho_S^h$ denote the nucleon scalar density and the scalar density
for a meson of type $h$, respectively. The scalar density of mesons
$h$ is evaluated in the independent-particle approximation, whereas the
nucleon scalar density $\rho_S$ has to be determined in a suitable
model with interacting degrees-of-freedom in order to match our
knowledge on the nuclear EoS at low temperature and finite density.
A proper (and widely used) approach is the non-linear
$\sigma-\omega$ model for nuclear matter which is used here with different parameter sets (NL1,NL2,NL3). \\

The main idea in Ref. \cite{Cassing:2015owa} is to consider effective
masses for the dressed quarks in the Schwinger formula
(\ref{schwinger}) for the string decay in a hot and dense
medium. The effective quark masses can be expressed in terms of a
scalar coupling to the quark condensate $\langle \bar q q \rangle$
in first order as follows:
\begin{equation}   \label{mss}
	m_s^* = m_s^0 + (m_s^v-m_s^0)\left| \frac{<\bar{q}q>}{<\bar{q}q>_V}
	\right| ,  m_q^* = m_q^0 + (m_q^v-m_q^0) \left| \frac{<\bar{q}q>}{<\bar{q}q>_V}
	\right| ,
\end{equation} using $ m_s^0 \approx$ 100 MeV and $m_q^0 \approx 7$
MeV for the bare quark masses.

\begin{figure}[h!]
	\centering
	\includegraphics[width=0.71\textwidth]{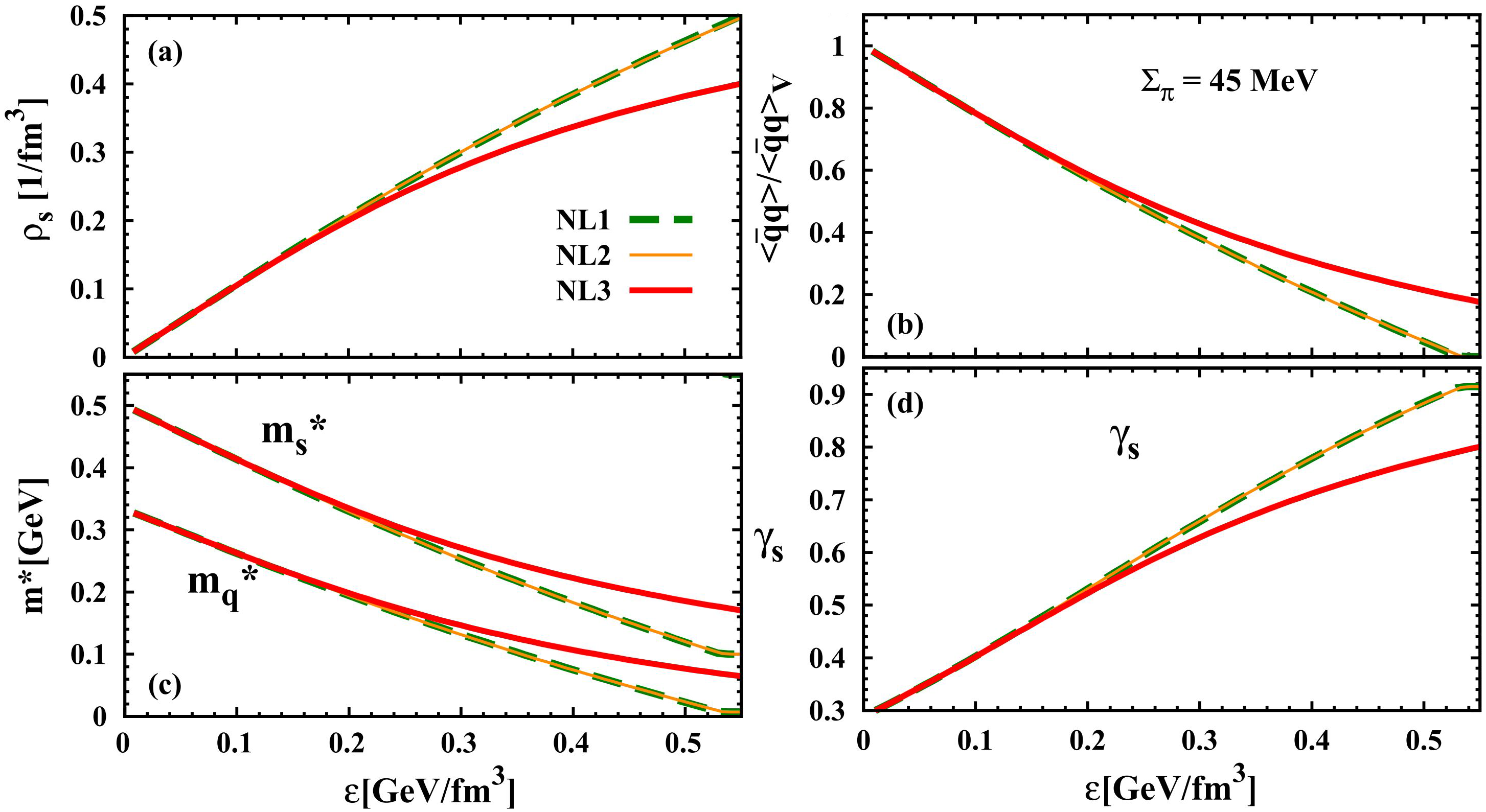}
	\vspace{-0.2cm}\caption{(Color online) The nucleon scalar density $\rho_S$ (a), the ratio between the scalar quark condensate and its value in the vacuum  $\langle \bar q q \rangle/\langle \bar q q \rangle_V$  (b), the light and strange quark effective masses $m^*_q, m^*_s$  (c), and the production probability of massive $s\bar s$ relative to light flavor production $\gamma_s$  (d) as a function of the energy density $\epsilon$ for the parameter sets NL3 (red solid lines), NL2 (thin orange lines)  and NL1 (dashed green lines) at $T=0$ and with $\Sigma_\pi$=45\, MeV.\vspace{-0.3cm}}
	\label{plots_EOS}
\end{figure}

\section{Application to nucleus-nucleus collisions and discussions}
\label{AA}

In this section we first study the excitation function of the particle
ratios $K^+/\pi^+$, $K^-/\pi^-$ and $(\Lambda+\Sigma^0)/\pi$ at
midrapidity from 5\% central Au+Au collisions. 
As already described in Ref. \cite{Cassing:2015owa}, the inclusion of CSR
in PHSD is responsible for the strong strangeness enhancement at AGS
and low SPS energies. The experimental observations of the ratios
$K^+/\pi^+$ and $(\Lambda+\Sigma^0)/\pi$ show the well-known "horn"
structure, which is reproduced by the PHSD calculations with CSR. In
fact, CSR gives rise to a steep increase of these ratios at energies
lower than $\sqrt{s_{NN}}\approx 7\,$GeV, while the drop at larger
energies is associated to the appearance of a deconfined partonic
medium. We point out that even adopting different parametrizations for the $\sigma-\omega$
model, we recover the same "horn" feature.

\begin{figure}[h!]
	\centering
	\includegraphics[width=1.\textwidth]{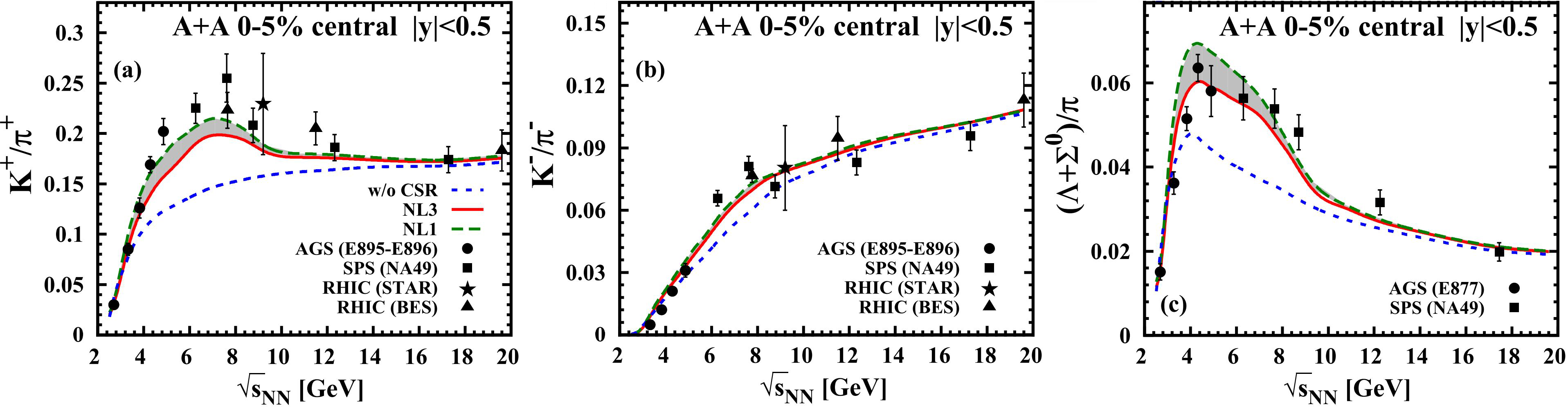}
	\vspace{-0.6cm}\caption{(Color online) The ratios $K^+/\pi^+$,
		$K^-/\pi^-$ and $(\Lambda+\Sigma^0)/\pi$ at midrapidity from 5\%
		central Au+Au collisions as a function of the invariant energy
		$\sqrt{s_{NN}}$ up to the top SPS energy in comparison to the
		experimental data from \cite{exp4}. The grey shaded area
		represents the results from PHSD including CSR taking into account
		the uncertainty from the parameters of the $\sigma-\omega$-model for
		the EoS.}
	\label{horn}
\end{figure}

Secondly, we analyze the dependence of the strange to
non-strange particle ratios on the size of the colliding system (cf.
also  Ref. \cite{Cleymans}). The inclusion of  CSR gives a strangeness enhancement
also in case of smaller system size with respect to Au+Au collisions
and this holds for all three particle ratios. In fact, when
considering central collisions, a sizeable volume of the system is
affected by the partial restoration of chiral symmetry even in case
of light ions. 
Concerning the "horn" structure in the $K^+/\pi^+$
ratio, we notice that the peak of the excitation function becomes
less pronounced in case of Ca+Ca and it disappears completely in
case of C+C collisions. With decreasing system size  the low energy
rise of the excitation functions becomes less pronounced. We can see
also that the peak for Ca+Ca is shifted to larger energies with
respect to the Au+Au case. Differently from the $K^+/\pi^+$, the
$(\Lambda+\Sigma^0)/\pi$ ratio preserves the same structure for all
three colliding systems.

\begin{figure}[h!]
	\centering
	\includegraphics[width=1.\textwidth]{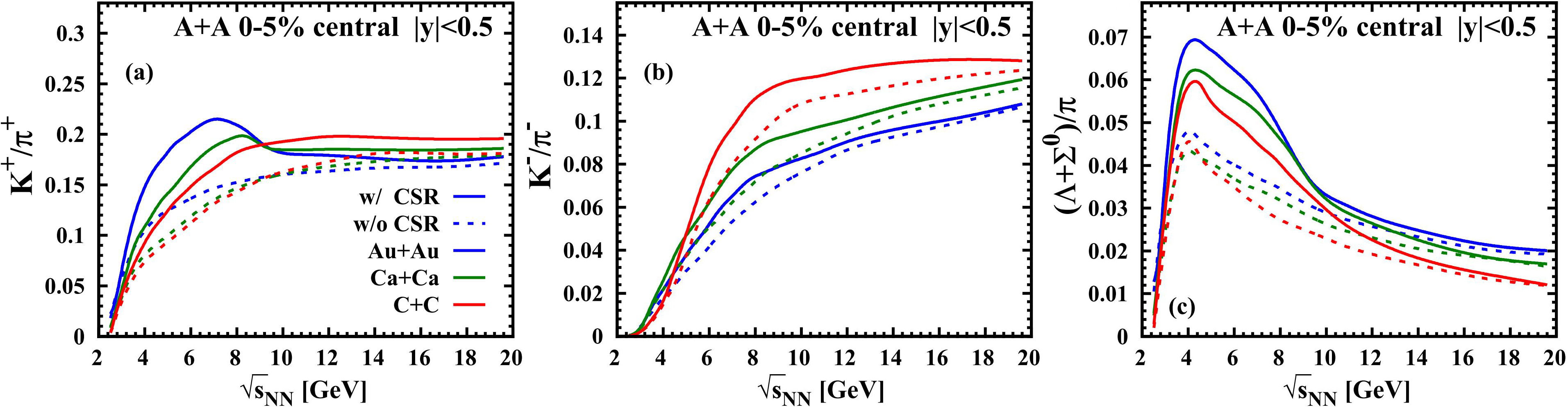}
	\vspace{-0.6cm}\caption{(Color online) The ratios $K^+/\pi^+$,
		$K^-/\pi^-$ and $(\Lambda+\Sigma^0)/\pi$ at midrapidity from 5\%
		central symmetric A+A collisions as a function of the invariant
		energy $\sqrt{s_{NN}}$. The solid lines show the results from PHSD
		including CSR with NL1 parameters, the dashed lines show the result
		from PHSD without CSR. The blue lines refer to Au+Au collisions, the
		green lines to Ca+Ca collisions and the red lines to C+C
		collisions.\vspace{-0.3cm}}
	\label{horn_systems}
\end{figure}

\begin{figure}[h!]
	\centering
	\includegraphics[width=1.\linewidth]{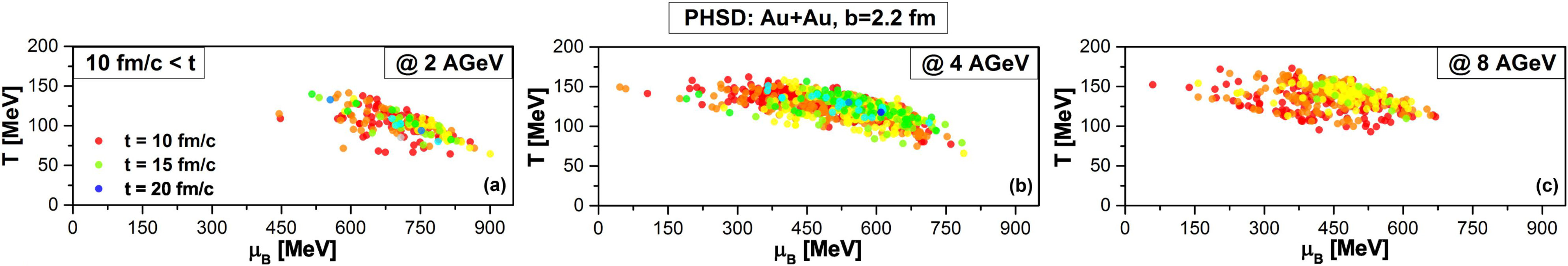}
	\vspace{-0.6cm}\caption{(Color online) Occupancy of the phase diagram for hadronic matter in Au+Au collisions at different beam energies from 2 to 8 AGeV for
		times $t > $ 10 fm/c. Each point belongs
		to a cell where strange quarks were produced. The color of the points indicates the time of the events within some varying interval.\vspace{-0.5cm}}
	\label{pic:pd_vs_energy}
\end{figure}

Finally, we study which parts of the phase diagram
in the ($T, \mu_B$)-plane are probed by heavy-ion collisions with
special focus on the strangeness production. One sees in Fig. \ref{pic:pd_vs_energy} that the probed region shifts to
larger baryon chemical potentials and smaller temperatures when
lowering the beam energy. In addition, it becomes apparent from Fig.
\ref{pic:pd_vs_energy} that it will be very hard to identify a
critical point in the ($T, \mu_B$)-plane experimentally since the
spread in $T$ and $\mu_B$ is very large at all bombarding energies
of interest.

\section*{Acknowledgements}
The authors acknowledge the support by BMBF, HIC for FAIR, H-QM and the HGS-HIRe for FAIR.
The computational resources were provided by the LOEWE-CSC.








\end{document}